\documentclass[11pt]{iopart}

\usepackage{iopams}
\usepackage{graphicx}
\usepackage{bm}

\graphicspath{{./Figures/}}

\newcommand{\qed}{\opensquare}

\newtheorem{lemma}{Lemma}

\begin{document}

\title[Lyapunov plateaus]{Families of piecewise linear maps with constant Lyapunov exponent}

\author{V. Botella-Soler}
\address{Departament de F\'{\i}sica Te\`{o}rica and IFIC,
Universitat de Val\`{e}ncia-CSIC, 46100-Burjassot,
Val\`{e}ncia, Spain}
\ead{vicente.botella@uv.es}
\author{J.A. Oteo}
\address{Departament de F\'{\i}sica Te\`{o}rica,
Universitat de Val\`{e}ncia, 46100-Burjassot,
Val\`{e}ncia, Spain}
\ead{oteo@uv.es}

\author{J. Ros}
\address{Departament de F\'{\i}sica Te\`{o}rica and IFIC,
Universitat de Val\`{e}ncia-CSIC, 46100-Burjassot,
Val\`{e}ncia, Spain}
\ead{jose.ros@uv.es}

\author{P. Glendinning}
\address{Centre for Interdisciplinary Computational and
Dynamical Analysis (CICADA) and School of Mathematics,
University of Manchester, Oxford Road, Manchester M13 9PL,
U.K.}
\ead{p.a.glendinning@manchester.ac.uk}

\begin{abstract}
We consider families of piecewise linear maps in which the moduli of the two slopes take different values. In some parameter regions, despite the variations in the dynamics, the Lyapunov exponent and the topological entropy remain constant. We provide numerical evidence of this fact and we prove it analytically for some special cases. The mechanism is very different from that of the logistic map and we conjecture that the Lyapunov plateaus reflect arithmetic relations between the slopes.

\end{abstract}

\vspace{2pc}


\section{Introduction}

Piecewise linear maps with a single discontinuity have been studied from many points of view. Renyi introduced
the $\beta-$transformations (maps of the form $\beta x$ (mod 1), $\beta >1$) to study invariant measures and
properties of Diophantine approximations, and the idea was taken up by Parry amongst others \cite{parry1960thebeta}. This led to a body
of work on arithmetic dynamics where arithmetic properties of the slopes (e.g. being Pisot numbers) lead to
special cases or transition points \cite{Sid}. Other approaches to these maps have been recently developed by
Dajani et al. \cite{dajani} and G\'ora \cite{gora2007, gora2009}. During the 1980s, piecewise monotonic maps 
with a single discontinuity were used to
model global bifurcations, leading to a greater understanding of the dynamics through kneading theory
and renormalization (or induced maps) \cite{MT, gambaudo1986new}. In the last few years there has been a renewed interest due to applications
in non-smooth bifurcation theory. Avrutin et al. have described many features of the piecewise linear cases from
this point of view \cite{avrutin2006,avrutin2008,ASB2006} and maps with constant slope have been derived as models of nonsmooth bifurcations of flows \cite{GKN, FG}.

In this paper we develop a theoretical approach to understand an observation made
in \cite{botella2009dynamics}. The maps studied are piecewise
linear with one increasing branch and one decreasing branch. For fixed values of these slopes there is a family
of maps having an invariant interval and parametrized by the difference in the values of the maps at the
discontinuity. The numerical experiments of \cite{botella2009dynamics}
suggest that there are values of the slopes such that the Lyapunov exponent (and, as we will show, the topological entropy)
of the map is constant over a range of choices for the values at the discontinuity. The reason for this turns out
to be a surprising robustness in the structure of the associated invariant measures, but there are still open
questions about precisely which slopes admit such plateaus. The answer appears to be arithmetic, involving
either Pisot numbers or, at the very least, more general algebraic integers.

Before describing the results in detail it is worth noting that it is quite easy to construct families of maps
with constant Lyapunov exponents and/or constant topological entropy, and so we will mention a few of these
constructions here to
contrast with the explanation for the phenomenon in this family. Our aim in the next few paragraphs is to show
that the results presented here are truly unexpected and cannot be understood by trivial mechanisms.

The most simple way to ensure that entropy and Lyapunov exponents are constant in a family is that all
elements of the family are smoothly conjugate to some map, $F:I\to I$, say. That is, there is a continuous
family of diffeomorphisms $q_\mu:I\to I$, parametrized by $\mu$, and the family of maps considered
is simply $G_\mu = q^{-1}_\mu\circ F\circ q_\mu$. The smooth conjugation of all elements in the family
$G_\mu$ to $F$ means that provided $F$ has positive entropy and well-defined Lyapunov exponent, then
all the functions $G_\mu$ also have the same values of these quantities.

The $\beta-$transformations themselves provide another example. Since the slope is always $\beta$ (where defined)
the Lyapunov exponent is trivially $\log\beta$, and so all $\beta-$transformations with the same value of $\beta$ have the
same Lyapunov exponent. Once again, our examples do not fall into this category, though there is a sense in which our
maps are closely related to the family of maps
\begin{equation}\label{eq:beta1}
P_{\mu,s}(x)=\left\{\begin{array}{ll} 1+sx &{\rm if}~x<0,\\ 1+\mu-sx &{\rm if}~x>0,\end{array}\right.
\end{equation}
with $s>1$. Provided the parameters are chosen so that an interval is mapped to itself then the entropy and
the Lyapunov exponent equal $\log s$. Milnor and Thurston prove (\cite{MT}, see also Glendinning \cite{G}
for an explicit account in this case) that if a piecewise monotonic
map $f$ with a single discontinuity and one branch increasing with the other branch decreasing has positive entropy then
it is semi-conjugate to one of the maps $P_{\mu ,s}$ and has entropy $\log s$, i.e. there is a monotonic
continuous but not necessarily invertible map $q$ such that $P_{\mu,s}\circ q=q\circ f$.
Since all the maps we consider have the properties of such $f$ they are each semi-conjugate to $P_{\mu,s}$ for some
$\mu$ and $s$. The surprise is that over ranges of parameters the families are semi-conjugate to maps
\emph{with the same value of s} even though the dynamics is changing, i.e. the values of $\mu$ is not constant.

One further way of creating families with constant entropy is a mechanism seen in the logistic (quadratic) map.
In this case the non-wandering set, at certain values of the parameters, can be decomposed into two parts:
a fixed part, on which the dynamics has positive entropy, and a second component whose dynamics changes with
the parameters but whose entropy is less than that of the fixed part, which is therefore the entropy of the map for all
the relevant parameter values. This is the case (for example) in the period three window, and more generally when
induced maps can be defined for higher iterates of the map. In the period three window there are
three sub-intervals (one containing the turning point) that are permuted by the map and mapped into themselves under
the third iterate. Between these three intervals there lies an invariant set with entropy $\log \frac{1+\sqrt{5}}{2}$, and
the maximum entropy of the third iterate is $\frac{1}{3}\log 2$, which is less than the entropy of the constant
part of the non-wandering sets. Glendinning and Hall \cite{GH} explore this, and the related kneading theory, 
for piecewise increasing maps. A description of topological entropy for these maps is given in \cite{kopf2000}.

The point we are making is that whilst there are a number of simple ways of finding families of maps with
constant entropy and Lyapunov exponent, the example presented here does not fit into these models. Indeed, another
way of looking at our results is that we have a two-parameter family of maps, and that we are interested in contours of
constant topological entropy. If these contours were in standard position there would be no reason to expect there
to be a simple choice of parameters such that a part of a contour is a line parallel to a parameter axis.
However, in our case this does happen, and we believe the reasons reflect arithmetic resonances in the equations.

In this paper, we restrict ourselves to showing that the phenomenon described really does happen. This will involve
constructing invariant measures for the maps and using these to describe the Lyapunov exponents. As such, we
provide further explicit examples of the type described more generally by G\'ora \cite{gora2009}. We will also use kneading
theory to prove the constancy of the topological entropy of the maps.

In \cite{botella2009dynamics} we investigated the properties of the map introduced by Varley, Gradwell and Hassell (VGH map) in \cite{varley1973insect} for the study of insect populations. The map has the following form
\begin{equation} \label{eq:Varley1}
y_{n+1}=\left\{
\begin{tabular}{ll}
      $ry_n,$ & $y\le c$ ,\\
      $ry_n^{1-b},$ & $y>c.$
     \end{tabular}
     \right.
\end{equation}
The map (\ref{eq:Varley1}) has an exact linearization in terms of the new parameters
\begin{equation}\label{eq:xi}
z\equiv 2\log(y)/\log(r), \quad \xi \equiv 2\log(c)/\log(r),
\end{equation}
with $y\ne 0$ and $r>1$,  that turns it into a piecewise linear map

\begin{equation} \label{eq:zlog}
z_{n+1}=\left\{
\begin{tabular}{ll}
      $z_n + 2,$ & $z_n\le \xi$, \\
      $(1-b)z_n +2,$ & $z_n>\xi$.
     \end{tabular}
     \right.
\end{equation}
A further change of variable ($x=z-\xi$) leaves the map in the form
\begin{equation} \label{eq:Tmap}
x_{n+1}=T(x_n)=\left\{
\begin{tabular}{ll}
      $x_n + 2,$ & $x_n\le 0$, \\
      $-s x_n +\beta,$ & $x_n>0$,
     \end{tabular}
     \right.
\end{equation}
with $\beta=2-b\xi$ and $s=b-1$. 

\begin{figure}[h]
\begin{center}
\includegraphics[width=8cm]{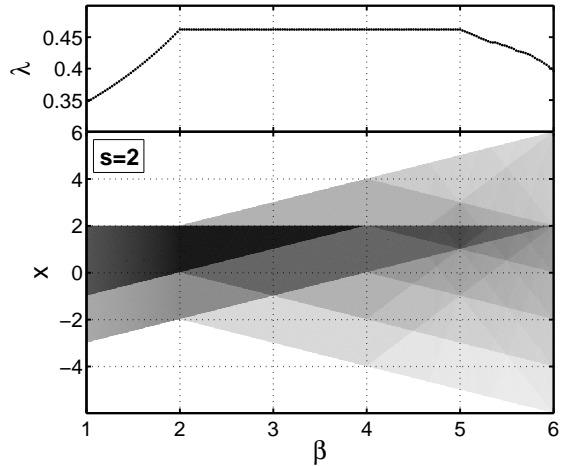}
\caption{\label{fig:LyapBif} Bifurcation diagram (bottom panel) and numerical estimation of the Lyapunov exponent (upper panel) as a function of $\beta$ for $s=2$. The Lyapunov exponent appears to be constant in the range $2\leqslant\beta\leqslant5$.}
\end{center}
\end{figure}

As we showed in \cite{botella2009dynamics} the map $T(x)$ is chaotic for $s>1$. In the rest of the paper we consider only this chaotic regime and restrict our attention to the invariant set which is $[(1-s)\beta,\beta]$ if $\beta\ge 2$, and $[\beta-2s,2]$ if $\beta<2$. In figure~\ref{fig:LyapBif} the bifurcation diagram as a function of parameter $\beta$ and the corresponding Lyapunov exponent are shown for the particular case $s=2$. The dynamics is chaotic with positive Lyapunov exponent. Interestingly, the numerically computed Lyapunov exponent shows a \textit{plateau} in the parameter range $2\le\beta\le5$ on which the Lyapunov exponent appears to be constant. The goal of this paper is to show that this is a real phenomenon and not a mere numerical artifact. To do so we calculate the Lyapunov exponent analytically which in this case reduces to finding a solution of the Perron-Frobenius equation for the invariant density. Moreover, we also prove that there is a corresponding plateau for the topological entropy; numerical experiments suggest that this coincides with the plateau of Lyapunov exponents.

By kneading theory arguments it is possible to show that the dynamics of the map change across the plateau. And also, that none of the simple mechanisms causing Lyapunov exponent plateaus that were described above apply in our case. 

The structure of the paper is as follows. In Section \ref{sec:Lyap} we describe the Lyapunov exponent plateaus. In Section \ref{sec:LyapPF} we explain how to calculate them analytically via the Perron-Frobenius equation. We illustrate this method with two simple examples and leave the complete proofs to \ref{app:proofs2} and \ref{app:proofsGR}. Similarly, in Section \ref{sec:kneading} we show that the Lyapunov exponent plateaus are accompanied by plateaus of the topological entropy and we also prove these making use of kneading theory. In Section \ref{sec:disc} we summarize our results and propose a list of open challenges to stimulate further work.

\section{Lyapunov exponent}
\label{sec:Lyap}

The Lyapunov exponent measures the average exponential rate of divergence of two initially close orbits of a dynamical system, with a positive exponent indicating chaotic behaviour. In the case of a discrete map $f$, the Lyapunov exponent of $x_0$ is

\begin{equation}\label{eq:Lyap1}
\lambda(x_0, f)=\lim_{n\to \infty}\left\{
\frac{1}{n}\sum_{k=0}^{n-1} \ln|f'(x_k)|  \right\}
\end{equation}
provided the limit exists.

If $\mu$ is an invariant measure for $f$ then the Lyapunov exponent of $f$ with respect to $\mu$ is

\begin{equation}\label{eq:Lyap2}
\lambda_\mu=\int \ln|f'(x)|d\mu(x),
\end{equation}
and for $\mu$-almost all points this equals (\ref{eq:Lyap1}). For our map, $\ln|f'(x)|=0$ if $x<0$ so (\ref{eq:Lyap1}) becomes

\begin{equation}\label{eq:Lyap}
\lambda=\ln|s|\lim_{n\to \infty}\left\{
\frac{1}{n}\sum_{k=0}^{n-1} \theta(x_k)  \right\},
\end{equation}
where $\theta(x)$ stands for the Heaviside step function. Similarly, equation (\ref{eq:Lyap2}) becomes

\begin{equation}\label{eq:Lyapmu}
\lambda_\mu=\ln|s| \int_{0}^{\infty} d\mu(x).
\end{equation}

It is clear from expressions (\ref{eq:Lyap}) and (\ref{eq:Lyapmu}) that what determines the exact value of the Lyapunov exponent is the fraction of time (or number of iterations) an orbit spends, on average, in the region $x>0$. Therefore, proving the constancy of the Lyapunov exponent in the plateau reduces to proving the constancy of $\mu([0,\infty])$.

\begin{figure}[h]
\begin{center}
\includegraphics[width=8cm]{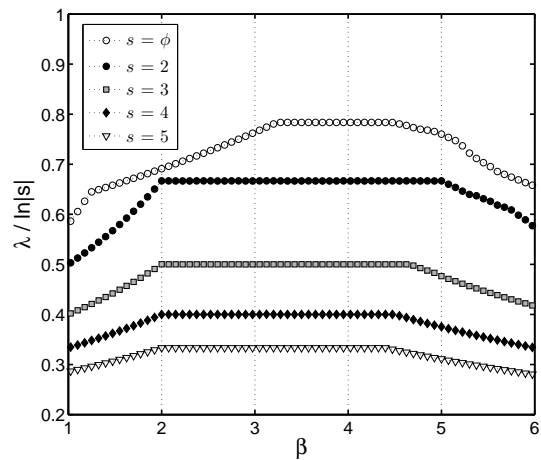}
\caption{\label{fig:Lyaps} Lyapunov exponents as a function of $\beta$ for different values of $s\in\mathbb{N}$ and $s=\phi\equiv\frac{1+\sqrt{5}}{2}$.}
\end{center}
\end{figure}

\begin{figure}[h]
\begin{center}
\includegraphics[width=8cm]{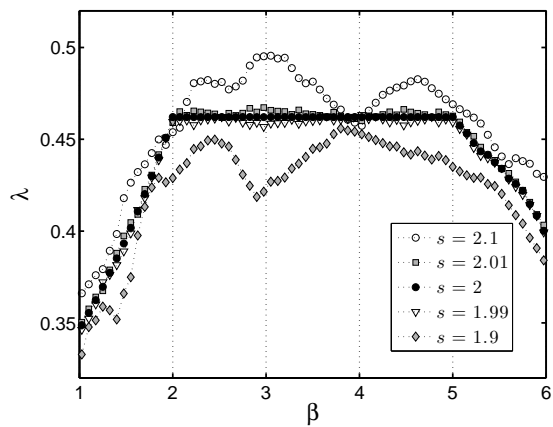}
\caption{\label{fig:Lyaps2} Lyapunov exponents as a function of $\beta$ for different values of $s$ close to the integer value $s=2$.}
\end{center}
\end{figure}

\begin{figure}[h]
\begin{center}
\includegraphics[width=8cm]{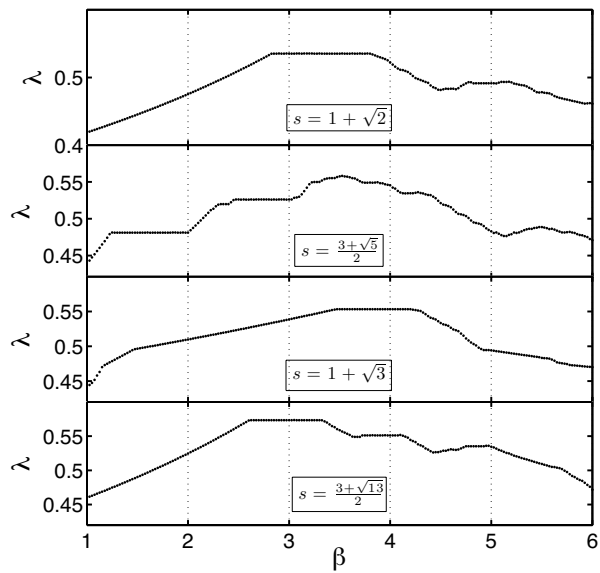}
\caption{\label{fig:LyapsPisot} Lyapunov exponents as a function of $\beta$ for the case of $s$ taking the value of different quadratic Pisot numbers.}
\end{center}
\end{figure}

Figure~\ref{fig:Lyaps} shows different plateaus for different values of $s$. Apparently, only some special values of $s$ show a plateau of constant value of the Lyapunov exponent.  In fact, as figure~\ref{fig:Lyaps2} illustrates, this plateau disappears if we vary the value of $s$ slightly. When $s\in\mathbb{N}$ the Lyapunov exponent remains constant in a range of $\beta$ of varying length (depending on the specific value of $s$) but starting in all cases at $\beta=2$. When $s$ takes other algebraic integer values, like Pisot numbers (including the golden ratio, the silver ratio, etc.), plateaus of constant Lyapunov exponent can also be observed (figure~\ref{fig:LyapsPisot}).

We have been able to prove that these plateaus are not a numerical artifact by determining analytically the Lyapunov exponent for two particular values of $s$ ($s=2$ and $s=\phi\equiv\frac{1+\sqrt{5}}{2}$). Taking into account (\ref{eq:Lyapmu}) the determination of the Lyapunov exponent is reduced to finding the natural invariant measure $\mu(x)$ of the attractor. To do this, it is necessary to solve the Perron-Frobenius equation for the invariant density. This process is explained in the next section. 
\section{Lyapunov exponents and the Perron-Frobenius equation}
\label{sec:LyapPF}

In general, note that if an invariant measure $\mu$ exists with a density $\rho(x)$ such that

\begin{equation}
\mu([a,b])=\int_a^b d\mu(x)=\int_a^b\rho(x)dx,
\end{equation}
then this density satisfies the Perron-Frobenius equation

\begin{equation}
\rho (x) =\sum_{\{y|T(y)=x\}}\frac{\rho (y)}{|T^\prime (y)|}.
\end{equation}
This is a functional equation for the density and there is no standard procedure to solve it. However, for some parameter ranges we have been able to pose an ansatz for $\rho(x)$ that we can prove satisfies the Perron-Frobenius equation and is therefore a solution. In particular, we find a partition such that by assuming the invariant density is piecewise constant, we are able to turn de Perron-Frobenius equation into a simple system of algebraic equations.

For the case $\beta>2$ the Perron-Frobenius equation can be written as

\begin{equation}\label{eq:PF}
 \rho (x) =\left\{\begin{array}{ll}
 \frac{1}{s}\rho (\frac{1}{s}(\beta -x)) & {\rm if}~x\in ((1-s)\beta , (1-s)\beta+2),\\
 \rho (x-2)+ \frac{1}{s}\rho (\frac{1}{s}(\beta -x)) & {\rm if}~x\in ((1-s)\beta+2 , 2),\\
\frac{1}{s}\rho (\frac{1}{s}(\beta -x)) & {\rm if}~x\in (2 , \beta).
 \end{array}\right.
\end{equation}

We will now illustrate this general method by solving two simple examples for particular values of the parameters.

\subsection{Two simple examples for $s=2$.}

\subsubsection{$\beta=2$.}

If $\beta =2$ then the map is continuous and one of the regions in (\ref{eq:PF}) is trivial, so

\begin{equation}\label{eq:PF2}
 \rho (x) =\left\{\begin{array}{ll}
 \frac{1}{2}\rho (\frac{1}{2}(2 -x)) & {\rm if}~x\in (-2 , 0),\\
 \rho (x-2)+ \frac{1}{2}\rho (\frac{1}{2}(2 -x)) & {\rm if}~x\in (0 , 2).
 \end{array}\right.
\end{equation}
Considering now the preimages of the intervals $I_1=(-2,0)$ and $I_2=(0,2)$

\begin{equation}
T^{-1}(I_1)=(1,2),
\quad
T^{-1}(I_2)=(-2,0)\cup(0,1),
\end{equation}
and assuming the invariant density is piecewise constant as an ansatz,
\begin{equation}
 \rho (x) =\left\{\begin{array}{ll}
 A & {\rm if}~x\in (-2 , 0),\\
 B  & {\rm if}~x\in (0, 2),\\
  \end{array} \right.
\end{equation}
the Perron-Frobenius equation (\ref{eq:PF2}) gives
\begin{equation}
  A =\frac{1}{2}B,\quad
 B= A+\frac{1}{2}B.
\end{equation}
It is easy to verify the simple piecewise constant solution
\begin{equation}\label{eq:PF2sol}
 \rho (x) =\left\{\begin{array}{ll}
 \frac{1}{6} & {\rm if}~x\in (-2 , 0),\\
 \frac{1}{3} & {\rm if}~x\in (0 , 2),
 \end{array}\right.
\end{equation}
where the normalization has been chosen so that $\int_{-2}^{2}\rho (x)dx=2(A+B)=1$. Hence, the invariant measure for $x>0$ reduces to
\begin{equation}
\mu ([0,2])=\int_0^2\frac{1}{3}dx=\frac{2}{3},
\end{equation}
and (\ref{eq:Lyapmu}) gives
\begin{equation}
\lambda = \frac{2}{3}\ln 2.
\end{equation}

\subsubsection{$\beta=4$.}

The previous case was so easy that the solution could be presented with no explanation at all. The case
$\beta =4$ is a little more complicated and serves to illustrate how the complexity of the calculation
increases with more complicated dynamics. The Perron-Frobenius equation (\ref{eq:PF}) becomes
\begin{equation}\label{eq:PF4}
 \rho (x) =\left\{\begin{array}{ll}
 \frac{1}{2}\rho (\frac{1}{2}(4 -x)) & {\rm if}~x\in (-4 , -2),\\
 \rho (x-2)+ \frac{1}{2}\rho (\frac{1}{2}(4 -x)) & {\rm if}~x\in (-2 , 2),\\
 \frac{1}{2}\rho (\frac{1}{2}(4 -x)) & {\rm if}~x\in (2 , 4).
 \end{array} \right.
\end{equation}
We consider in this case the following intervals
\begin{equation}\begin{array}{rl}
I_1&=(-4,-2), \\
I_2&=(-2,0), \\
I_3&=(0,2), \\
I_4&=(2,4),
\end{array}\end{equation}
whose preimages are
\begin{equation}\begin{array}{rl}
T^{-1}(I_1) &=(3,4),\\
T^{-1}(I_2) &=(-4,-2)\cup(2,3),\\
T^{-1}(I_3) &= (-2,0)\cup(1,2),\\
T^{-1}(I_4) &= (0,1).
\end{array}\end{equation}
This suggests that we solve
(\ref{eq:PF4}) by posing the piecewise constant solution
\begin{equation}
 \rho (x) =\left\{\begin{array}{ll}
 A & {\rm if}~x\in (-4 , -2),\\
 B  & {\rm if}~x\in (-2 , 0),\\
 C  & {\rm if}~x\in (0 , 2),\\
D & {\rm if}~x\in (2 , 4),
 \end{array} \right.
\end{equation}
in which case (\ref{eq:PF4}) implies
\begin{equation}
A=\textstyle{\frac{1}{2}}D, \quad B=A+\textstyle{\frac{1}{2}}D ,\quad C=B+\textstyle{\frac{1}{2}}C, \quad D=\textstyle{\frac{1}{2}}C.
\end{equation}
The normalization requirement $\int \rho dx=1$ in this case reads
\begin{equation}
2(A+B+C+D)=1 .
\end{equation}
An elementary calculation shows that this is solved by
\begin{equation}
A=\textstyle{\frac{1}{18}}, \quad B=\textstyle{\frac{1}{9}}, \quad C= \textstyle{\frac{2}{9}},\quad D=\textstyle{\frac{1}{9}},
\end{equation}
and so the invariant measure in $x>0$ is
\begin{equation}
\mu ([0,4])=\int_0^2\frac{2}{9}dx+\int_2^4\frac{1}{9}dx=\frac{4}{9}+\frac{2}{9}=\frac{2}{3},
\end{equation}
and the Lyapunov exponent is
\begin{equation}
\lambda = \frac{2}{3}\ln 2,
\end{equation}
which is equal to the Lyapunov exponent of the previous example.

\subsection{Analytical proofs of two plateaus}

Following the same methodology as in the previous examples it is possible to solve the Perron-Frobenius equation as a function of $\beta$. This is possible mainly due to the robustness of the invariant density $\rho(x)$ with varying $\beta$. This allows us to find finite partitions of phase space on which the value of $\rho(x)$ in each interval remains constant although the size of the intervals can vary with $\beta$. The robustness of $\rho(x)$ can be observed in figure~\ref{fig:LyapBif} where the different shades of grey indicate different values of the invariant density. Between $\beta=2$ and $\beta=3$ for instance, we can see that the number of intervals remains constant and also does the density in each of the intervals. Moreover, the borders of the intervals as a function of $\beta$ are straight lines which further facilitates the partitioning of phase space and finding an appropriate ansatz for $\rho(x)$.

\begin{figure}[h]
\begin{center}
\includegraphics[width=8cm]{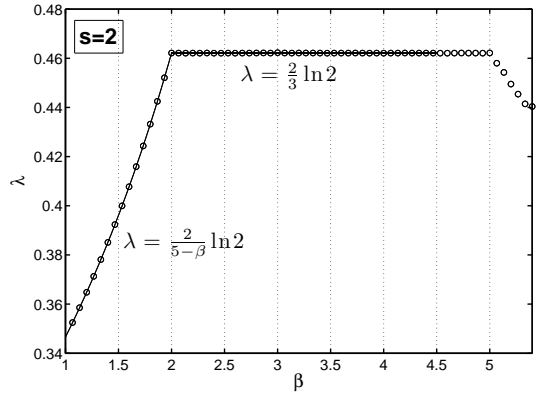}
\caption{\label{fig:Lyap_ana} Lyapunov exponent as a function of $\beta$ for $s=2$. The circles represent the numerical estimation of the Lyapunov exponent. The solid line stands for the analytical Lyapunov exponent calculated from the Perron-Frobenius equation.}
\end{center}
\end{figure}

\begin{figure}[h]
\begin{center}
\includegraphics[width=8cm]{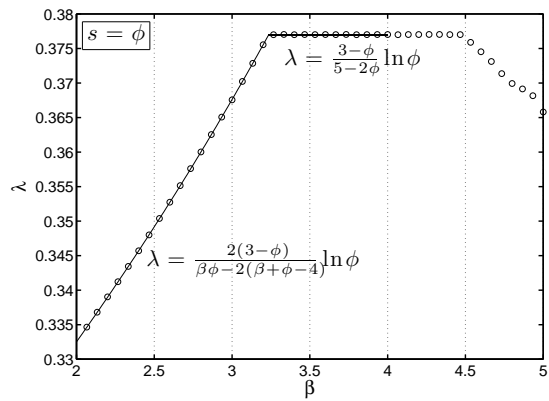}
\caption{\label{fig:Lyap_ana_GR} Lyapunov exponent as a function of $\beta$ for $s=\phi$. The circles represent the numerical estimation of the Lyapunov exponent. The solid line stands for the analytical Lyapunov exponent calculated from the Perron-Frobenius equation.}
\end{center}
\end{figure}

We have proved the existence of plateaus for the cases $s=2$ and $s=\phi$. However, we have not been able to prove them in its entirety because the structure of the invariant density gets increasingly complex as $\beta$ grows. The results are shown in figure~\ref{fig:Lyap_ana} and figure~\ref{fig:Lyap_ana_GR}. To be specific, the calculations detailed in \ref{app:proofs2} and \ref{app:proofsGR} prove the following lemma.

\begin{lemma}
Consider the maps defined by (\ref{eq:Tmap}) with $\beta\ge 2$. If $s=2$ and $2< \beta< \frac{9}{2}$ then the Lyapunov exponent is constant and equal to $\frac{2}{3}\ln 2$. If $s=\phi=\frac{1+\sqrt{5}}{2}$ and $2\phi<\beta<4$ then the Lyapunov exponent is a constant and equal to $\frac{3-\phi}{5-2\phi}\ln\phi$.
\end{lemma}

\section{Topological entropy and kneading theory}
\label{sec:kneading}

\begin{figure}
\begin{center}
\includegraphics[width=8cm]{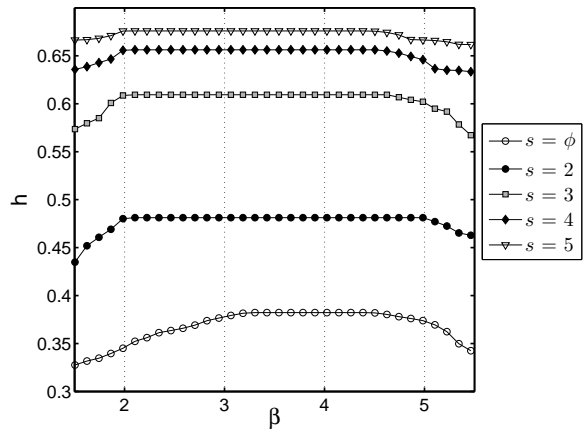}
\caption{\label{fig:entplat} Plots of numerically calculated topological entropy as a function
of $\beta$ using terms up to and including $t^{100}$ in (\ref{eq:zeros}) for different values of $s\in\mathbb{N}$ and $s=\phi=\frac{1+\sqrt{5}}{2}$. }
\end{center}
\end{figure}

\begin{figure}
\begin{center}
\includegraphics[width=8cm]{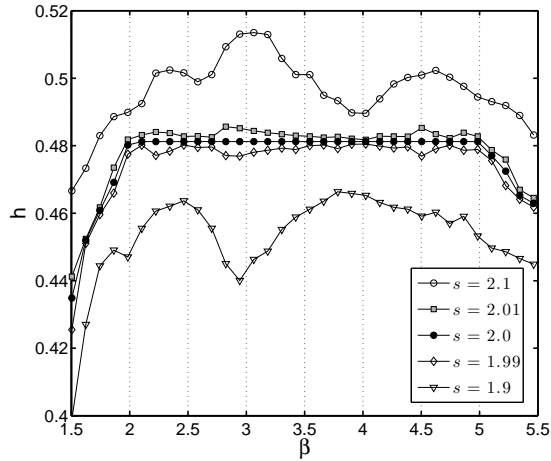}
\caption{\label{fig:entplat2} Plots of numerically calculated topological entropy as a function
of $\beta$ using terms up to and including $t^{100}$ in (\ref{eq:zeros}) for different values of $s$ close to the integer value $s=2$.}
\end{center}
\end{figure}

\begin{figure}
\begin{center}
\includegraphics[width=7cm]{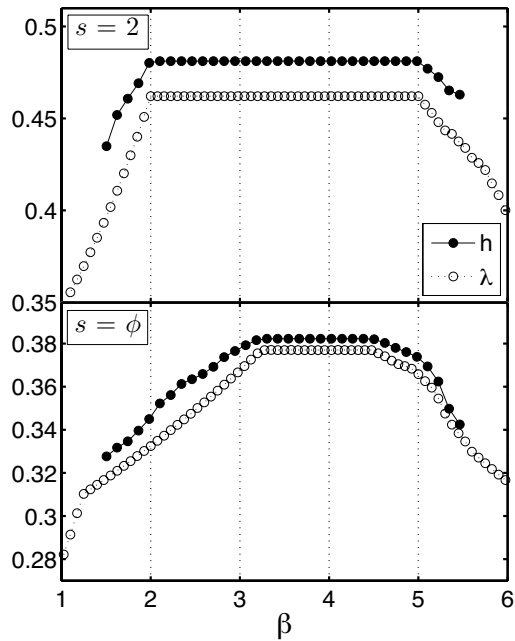}
\caption{\label{fig:lyapentplat} Plots of numerically calculated topological entropy and Lyapunov exponent as a function
of $\beta$.}
\end{center}
\end{figure}

Figures~\ref{fig:entplat} and \ref{fig:entplat2} shows the numerically calculated topological entropy $h$ of the maps
as a function of $\beta$ for different values of $s$. They present plateaus in precisely the same
positions as the Lyapunov exponent plateaus of figure~\ref{fig:Lyaps} as shown in figure~\ref{fig:lyapentplat}. Thus we conjecture that the
Lyapunov plateaus are also parameter intervals with constant topological entropy. This
assumption makes it possible to investigate other possible plateaus in more detail, but before
presenting the results we will sketch how the topological entropy is calculated.

Milnor and Thurston \cite{MT} developed kneading theory for unimodal maps, and noted that by
considering the kneading invariant of the map as a power series then if the map has positive
topological entropy this is equal to minus the logarithm of the smallest positive zero of the kneading invariant. Moreover, the kneading sequence (seen as power series evaluated
at this smallest zero of the kneading invariant) then provides a semi-conjugacy to a tent map
with slopes having constant absolute value equal to the reciprocal of the smallest zero. Glendinning and Hall \cite{GH}
showed how these results could be extended to Lorenz maps (piecewise increasing maps with a single discontinuity)
and Glendinning \cite{G} describes the theory for maps with a single discontinuity, one branch of which is
increasing and the other decreasing. Proofs of the results described here can be found in \cite{MT, G}.

Suppose that $f:[-a,b]\to [-a,b]$ with $a,b>0$, is a map with a single discontinuity at $x=0$, continuous and increasing on
$(-a,0)$ and continuous and decreasing on $(0,b)$. Then the standard symbolic description (e.g. \cite{CE})
assigns an address to each point $x\ne 0$ by $A(x)=1$ if $x>0$ and $A(x)=-1$ if $x<0$. The itinerary
of a point that is not a preimage of zero is then just the sequence
\begin{equation}
I(x)=A(x)A(f(x))A(f^2(x))A(f^3(x))\dots .
\end{equation}
Whilst the itinerary is easy to interpret, Milnor and Thurston \cite{MT} observed that the same information is contained
in a sequence of coordinates that monitor the slope of iterates of the map. It is natural to use a plus sign for
an increasing slope and a minus for decreasing slope, so let
\begin{equation}\label{eq:kncoord}
\theta_0 (x)=-A(x), \quad \theta_n(x)=-A(f^n(x))\theta_{n-1}(x),~n\ge 1
\end{equation}
for points that are not preimages of zero. Thus the sign of $\theta_n(x)$ is the sign of the slope of $f^{n+1}$
at $x$. Moreover, rather than sequences we can work with formal power series
\begin{equation}
P(x,t)=\sum_{k=0}^\infty \theta_k(x)t^k
\end{equation}
which is called the \emph{kneading sequence} of $x$. For points that are preimages of zero, two
kneading sequences (the \emph{lower} and \emph{upper} kneading sequences) can be defined:
\begin{equation}
P(x_-,t)=\lim_{y\uparrow 0}P(y,t),\quad   P(x_+,t)=\lim_{y\downarrow 0}P(y,t)
\end{equation}
where limits are taken through points that are not preimages of zero.

With the standard lexicographical order, ($\sum a_kt^k < \sum b_kt^k$ if $a_r=b_r$, $r=0,\dots ,j-1$,
and $a_j<b_j$) the sequences $P(x,t)$ are decreasing functions of $x$ and by looking at the difference
$P(x_-,t)-P(x_+,t)$ this function becomes monotonic and continuous as a function of $x$ if
$t=t^*$, where $t^*<1$ is the smallest positive zero of
\begin{equation}\label{eq:zeros}
P(0_-,t)-P(0_+,t)=0 .
\end{equation}
Such a value of $t^*$ always exists if the topological entropy of $f$ is positive and the entropy
actually equals $h=-\log t^*$ (see below).

The seminal result of Milnor and Thurston \cite{MT}, which carries over to discontinuous maps of the
kind considered here \cite{G}, is that if $f$ has positive topological entropy then the function
\begin{equation}\label{eq:h}
q(x)=\frac{P(x_\pm ,t^*)-P(-a_+,t^*)}{P(b_-,t^*)-P(-a_+,t^*)}
\end{equation}
is a semi-conjugacy (monotonic and continuous) from $f$ to a piecewise continuous map with
slopes having modulus $1/t^*$ and the topological entropy of $f$ is $h=-\log t^*$.

Details can be found in \cite{ MT, G}, but from the point of view of this paper the important point is that
the entropy can be calculated by looking at the zeroes of (\ref{eq:zeros}) by truncating the series
at order $n$. Keeping only polynomial terms up to $t^n$, the maximum error is $\frac{t^{n+1}}{1-t}$, and since $t\in [\frac{1}{2},1)$
(as the entropy of a two branch map of an interval into itself cannot be greater than $\log 2$)
this gets small as $n$ gets large. This is how figure~\ref{fig:entplat} and \ref{fig:entplat2} were calculated.

\begin{figure}
\begin{center}
\includegraphics[width=9cm]{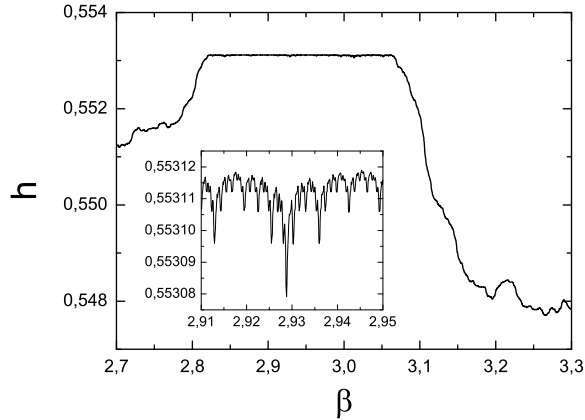}
\caption{\label{fig:noplat} Plots of numerically calculated topological entropy as a function
of $\beta$ with slopes 1.1 and -2.2 in \ref{eq:Tmap} using terms up to and including $t^{100}$ in (\ref{eq:zeros}).}
\end{center}
\end{figure}

Figures~\ref{fig:entplat} and \ref{fig:lyapentplat} provide strong evidence that the constant Lyapunov exponent plateaus
described above, and which we have proved exist if the slopes are $(1,-s)$ for a variety of choices
of $s$ (integers, some Pisot numbers) are concurrent with plateaus in the topological entropy
of the maps. This conjecture provides a way to attempt a further investigation into ambiguous
`almost plateaus' observed for the Lyapunov exponents in, for example, the case with slopes $(1.1,-2.2)$.
Iteration and root-finding algorithms make it relatively easy to compute roots of (\ref{eq:zeros})
to high accuracy. The entropy obtained for the case for slopes $(1.1,-2.2)$ is shown in figure~\ref{fig:noplat}. In this figure
the possibility of a plateau is apparent, as it was for the Lyapunov exponent. However,
a close up of parameter values in the apparent plateau demonstrates that there is actually a great deal
of structure in the variation of the topological entropy, and so we believe that this is \emph{not} a plateau
for the topological entropy of the maps. Assuming that the conjectured connection between plateaus in
the entropy and plateaus in the Lyapunov exponent holds, this implies that the Lyapunov exponent
plateaus only exist at interesting resonances of the form $(1,-s)$ and are not due to
resonances associated with the ratio of the slopes more generally.

We will prove the existence of plateaus for the topological entropy in two cases: $s=2$ and
$s=\phi$. The proof in both cases relies on the algebraic properties of the slope, and this
should provide a clue to the answer to the more general question about which values of $s$ have Lyapunov
plateaus. As with the Lyapunov exponents we do not prove the existence of the plateau for the largest possible
range of values of $\beta$ -- to consider the entire interval using our techniques would require the separate consideration
of more and more complicated cases precisely as with the Lyapunov plateaus -- but we do prove the existence of a
plateau. Recall that our numerical results appear to show that the Lyapunov plateaus and the plateaus of
topological entropy are equal.

\begin{lemma}
Consider the map $T$ of (\ref{eq:Tmap}) with $s=2$. If $2\le \beta \le 4$ then the topological entropy
of $T$ is $\log \left(\frac{1+\sqrt{5}}{2}\right)$.
\end{lemma}

\textit{Proof:}
We begin by considering the iterates of $x=0$ approached from above and below respectively, denoting these
by $0_+$ and $0_-$ respectively. If $\beta=2$ or $\beta =4$ then the iterates of $0_-$ and $0_+$ are in a
finite set which makes it straightforward to use these points to create a Markov partition for which
the entropy is straightforward to calculate (cf. \cite{BGMY}). So assume that $2<\beta <4$. Then
by direct calculation
\begin{equation}
\fl
T(0_+)=\beta >0,\quad T(\beta )=-\beta <0, \quad T(-\beta )=2-\beta <0, \quad T(2-\beta )=4-\beta >0;
\end{equation}
and
\begin{equation}
\fl
T(0_-)=2 >0,\quad T(2 )=\beta -4 <0, \quad T(\beta -4)=\beta -2 >0, \quad T(\beta -2)=4-\beta >0;
\end{equation}
so in particular
\begin{equation}\label{eq:iteq}
T^4(0_+)=T^4(0_-) .
\end{equation}
This makes it possible to compute the difference of the kneading polynomials (\ref{eq:zeros}) explicitly
without knowing the full details of each individual series and hence compute the zeros. To be specific: using
(\ref{eq:kncoord}) we find
\begin{equation}
P_+(t)=-1+t+t^2+t^3+t^4P(4-\beta ,t )
\end{equation}
and
\begin{equation}
P_-(t)=1-t-t^2+t^3+t^4P(4-\beta ,t)
\end{equation}
(strictly speaking the unknown polynomial is $P((4-\beta)_-,t)$) and hence using (\ref{eq:zeros}) the
entropy is minus the logarithm of the smallest positive zero of $1-t-t^2$.
\begin{flushright}
\qed
\end{flushright}

The proof for the plateau with $s$ being the golden mean is similar:

\begin{lemma}Consider the map $T$ of (\ref{eq:Tmap}) with $s=\phi\equiv (1+\sqrt{5})/2$.
If $2\phi\le \beta < 2\phi +1$ then the topological entropy
of $T$ is $-\log t^*$ where $t^*$ is the smallest positive zero of $1-t-t^3$.
\end{lemma}

\textit{Proof:}
Again, the end-point is easy to deal with separately so assume that $2\phi <\beta <2\phi +1$ and
write $\beta =2\phi +\epsilon$, $0<\epsilon <1$. Then by direct calculation using the relation
$\phi^2=1+\phi$ to simplify nonlinear terms in $\phi$:
\begin{eqnarray}
\fl
T(0_+)=2\phi +\epsilon >0,\quad T^2(0_+ )=-2-\epsilon (\phi -1) <0, \quad T^3(0_+ )=-\epsilon (\phi -1) <0,\nonumber\\
T^4(0_+)=2-\epsilon (\phi-1) >0, \quad T^5(0_+)=2\epsilon >0;
\end{eqnarray}
and
\begin{eqnarray}
T(0_-)=2 >0,\quad T^2(0_- )=\epsilon >0, \quad T^3(0_-)=2\phi -\epsilon (\phi -1) >0,\nonumber\\
T^4(0_-)=-2(1-\epsilon )<0, \quad T^5(0_-)=2\epsilon >0;
\end{eqnarray}
so
\begin{equation}\label{eq:iteqphi}
T^5(0_+)=T^5(0_-) .
\end{equation}
Thus
\begin{equation}
P_+(t)=-1+t+t^2+t^3-t^4-t^5P(2\epsilon ,t )
\end{equation}
and
\begin{equation}
P_-(t)=1-t+t^2-t^3-t^4-t^5P(2\epsilon ,t)
\end{equation}
where strictly speaking we should have checked that $2\epsilon$ is approached from the same side in
both cases. Hence using (\ref{eq:zeros}) the entropy is minus the logarithm of the smallest positive zero of
the difference of the two power series, i.e. of $1-t-t^3$.
\begin{flushright}
\qed
\end{flushright}

\begin{figure}
\begin{center}
\includegraphics[width=8cm,angle=270]{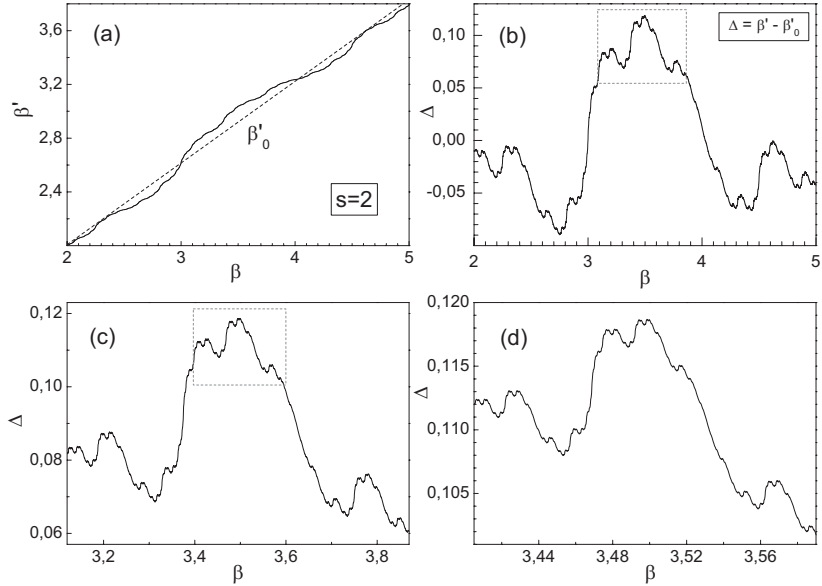}
\caption{\label{fig:selfsim2} (a) Plot of numerically obtained relation between $\beta'$ and $\beta$ for the case $s=2$. The dashed line corresponds to the linear regression $\beta'_0=0.79116+0.60802\beta$. (b) Detail of the fluctuation $\Delta=\beta' -\beta'_0$. (c) and (d) Successive zooms of the framed areas.}
\end{center}
\end{figure}

\begin{figure}
\begin{center}
\includegraphics[width=8cm,angle=270]{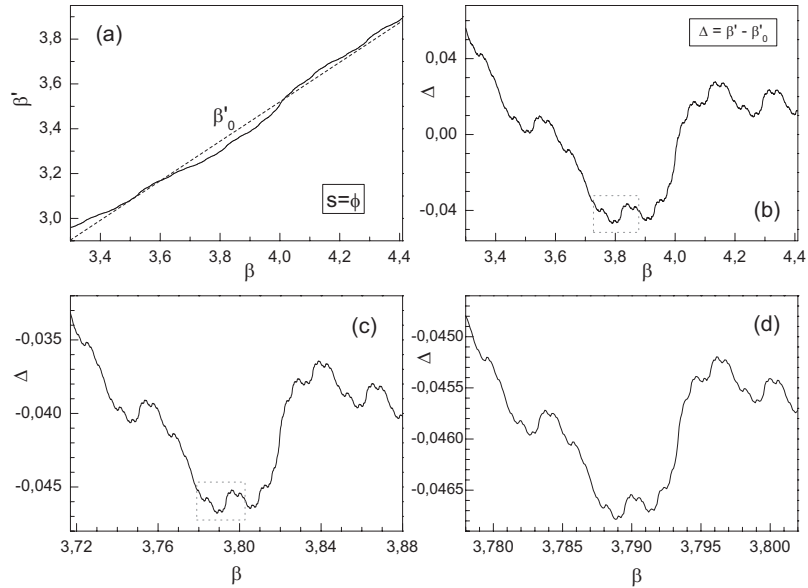}
\caption{\label{fig:selfsimGR} (a) Plot of numerically obtained relation between $\beta'$ and $\beta$ for the case $s=\phi$. The dashed line corresponds to the linear regression $\beta'_0=-0.00512+0.88136\beta$. (b) Detail of the fluctuation $\Delta=\beta' -\beta'_0$. (c) and (d) Successive zooms of the framed areas. }
\end{center}
\end{figure}

In the introduction we claimed that there are some fairly straightforward ways of creating families with
plateaus of Lyapunov exponents and topological entropy. Figure~\ref{fig:selfsim2} and figure~\ref{fig:selfsimGR}
provide evidence for our claim that our results do not fall into these `simple' categories.

By the kneading theory developed above, any two-branch map such as we are considering which has positive entropy
is semi-conjugate to a map with slopes having fixed modulus equal to $1/t^*$, where $t^*$ is the smallest positive zero
of the kneading difference (\ref{eq:zeros}). Thus each map in the plateau with entropy $-\log t^*$ is semi-conjugate
via the function $q$ defined in (\ref{eq:h}) to a map of the form
\begin{equation}\label{eq:beta2}
F_{\beta^\prime,s^*}(x)=\left\{\begin{array}{ll} 2+s^*x &{\rm if}~x<0,\\ \beta^\prime -s^*x &{\rm if}~x>0,\end{array}\right.
\end{equation}
where $s^*=1/t^*$ and we have rescaled (\ref{eq:beta1}) to match our parametrization for $T$. Thus for each $\beta$ in the plateau, $s^*$ is known and fixed, so there
is one parameter to be matched and the relationship between $\beta$ and the associated $\beta^\prime$ describes reveals the
way in which the dynamics of the maps in the plateaus change with the parameter $\beta$. Note that we believe that
in the plateaus the semi-conjugacy is actually a conjugacy, and the computer programme which generates
figures~\ref{fig:selfsim2} and~\ref{fig:selfsimGR} provides some evidence for this. The diagrams show the relationship
between the two parameters $\beta$ for $T$ and $\beta^\prime$ for $F$. It is computed by
calculating the kneading sequence of $0_+$ and $0_-$ for different values of $\beta$ in the plateau and then
uses a bisection algorithm based on the natural order for these sequences \cite{MT, G}
for each such $\beta$ to find the value of $\beta^\prime$ for which the
kneading sequence of $0_+$ equals the known sequence of $0_+$ for $\beta$. We then check that the sequences for
$0_-$ in both cases are the same to good accuracy which provides evidence that the two maps have the same
kneading series and that there is indeed a topological conjugacy (and not just a semi-conjugacy)
between the two maps.

The correspondence between $\beta$ and $\beta^\prime$ appears strictly monotonic, indicating that no two
maps in the plateau have the same dynamics. Moreover, there is a linear scaling and the variation about this
linear trend certainly appears very complicated if not fractal. This suggests that maps in the plateaus have a very rich
dynamic structure despite having the same Lyapunov exponents and topological entropy.

\section{Discussion}
\label{sec:disc}
In this paper we have proved that the plateaus in the Lyapunov exponent observed in
piecewise linear families of maps in \cite{botella2009dynamics} really do exist and have extended the
results to different slopes (using slopes $s=2$ and $s=\phi$) and we have also shown that the
plateau is also present in the topological entropy of the maps. We have provided strong numerical
evidence in figures~\ref{fig:selfsim2} and~\ref{fig:selfsimGR} that the dynamics varies in a
complicated fashion within the plateaus and, in particular, that the dynamics is certainly not constant within
the plateaus, which would have been an obvious but somewhat trivial way to obtain the plateaus. Numerical
experiments also suggest that this phenomenon is associated only with maps having one branch with slope equal to unity
and not with a simple resonance (where one slope is an integer multiple of the other for example).
Finally we have also obtained exact expressions for the Lyapunov exponent at the lower end of the plateaus
as it becomes different from the constant function on the plateaus.

The proof of the existence of the plateaus for the Lyapunov exponent uses the fact that we are able to calculate
invariant densities for the maps explicitly. These are piecewise constant on a finite union of intervals
similar to those calculated for some related examples by
G\'ora \cite{gora2009}, and the structure is robust within the plateaus: the number
of intervals remains constant over several sub-intervals of the plateau, as does the value of the density. Only
the lengths of the intervals vary. What makes this striking is that (as noted earlier) the dynamics
changes with changing parameter even though the structure of the invariant density does not. The number of
intervals on which the density changes at discrete values of the parameter, and these appear to accumulate
on the right hand end point of the plateau. 

The proof of the plateaus for the topological entropy of the maps suggests that one of the reasons for the existence
of the plateaus is a robust relation between iterates of the discontinuity approached from above and
from below. This is almost certainly the origin of the algebraic properties of those slopes for which we
observe plateaus: they are Pisot numbers. We
conjecture that if a family of maps has a plateau then the slope $s$ is a Pisot number. The relations between
the iterates of the discontinuity typified by (\ref{eq:iteq}) and (\ref{eq:iteqphi}) also change within the
plateaus, and we conjecture that these robust relationships are related to the robust structure of the
invariant densities.

\begin{figure}[h]
\begin{center}
\includegraphics[width=8cm]{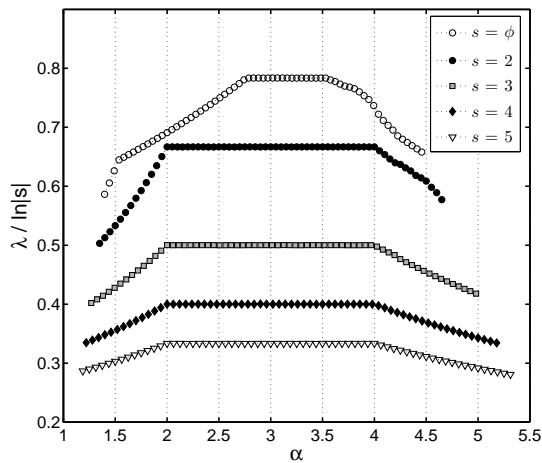}
\caption{\label{fig:LyapsAlpha} Lyapunov exponents as a function of $\alpha=\frac{s}{s+1}(\beta+\frac{2}{s})$ for different values of $s\in\mathbb{N}$ and $s=\phi\equiv\frac{1+\sqrt{5}}{2}$.}
\end{center}
\end{figure}

As well as the conjectures made above there remain a number of questions which would help illuminate this
phenomenon.
\begin{itemize}
\item Is it true that the plateaus only occur if one of the slopes equals one? If so, can the set of slopes
which produce plateaus be characterized completely?
\item What characterizes the end-points of the plateaus? Interestingly, it is possible to set all the plateaus for the case $s\in\mathbb{N}$ to the same length and in the same parameter range by parametrizing the map with $\alpha=\frac{s}{s+1}(\beta+\frac{2}{s})$ instead of $\beta$. As figure~\ref{fig:LyapsAlpha} shows, the Lyapunov plateaus for integer $s$ occur in the range $2<\alpha<4$. 
\item G\'ora \cite{gora2007,gora2009} provides a general expression for the invariant density of such maps. When
do these densities have the robust structure described above (what is their `bifurcation theory')? 
Our results also show that the existence of robust structure
continues to hold below the lower limit of the plateaus, so the robust structure with a finite set of 
intervals is not in itself enough to guarantee the existence of plateaus. How are the two cases different?
\item Do plateaus exist in other piecewise linear models with positive topological entropy and a single discontinuity?
\end{itemize}
This latter suggestion deserves a little more examination. The piecewise linear models described here have one increasing branch and one decreasing branch. There are two other possibilities (increasing-increasing and decreasing-decreasing)
and in the non-increasing case the bifurcation structures are interrelated \cite{GLT}. However, there is reason to believe
that neither of the other cases can have plateaus -- essentially because initial considerations suggest that the
only way to have $f^n(0_+)=f^n(0_-)$ is to have symmetric maps, and hence to be in the trivial cases we have effectively ruled
out here, where both branches have the same absolute value. If this is really the case then the coincidence here
becomes even more surprising, and lends credence to the suggestion that this phenomenon is an example of the interplay
of arithmetic with dynamics in similar ways to which certain other properties of the general $\beta-$transformations
are considered in the area of arithmetic dynamics (e.g. \cite{Sid}).

\ack{The authors thank Nikita Sidorov for useful conversations on this work.
PG is partially funded by EPSRC grant EP/E050441/1. VBS, JAO and JR are
partially supported by contract MICINN (AYA2010-22111-C03-02). VBS also
thanks Generalitat Valenciana for financial support.}

\appendix
\section{Lyapunov exponent plateau for $s=2$}
\label{app:proofs2}

\subsection{$2<\beta<\frac{9}{2}$}

The Perron-Frobenius equation (\ref{eq:PF}) in this case reads

\begin{equation}\label{eq:PFbeta2}
 \rho (x) =\left\{\begin{array}{ll}
 \frac{1}{2}\rho (\frac{1}{2}(\beta -x)) & {\rm if}~x\in (-\beta , -\beta+2),\\
 \rho (x-2)+ \frac{1}{2}\rho (\frac{1}{2}(\beta -x)) & {\rm if}~x\in (-\beta+2 , 2),\\
 \frac{1}{2}\rho (\frac{1}{2}(\beta -x)) & {\rm if}~x\in (2 , \beta).
 \end{array} \right.
\end{equation}
We start by considering the parameter range $2<\beta<5/2$. We will do the calculation for this case in detail. Consider nine intervals

\begin{equation}\begin{array}{rl}
I_1&=(-\beta,\beta-4), \\
I_2&=(\beta-4,-\beta+2), \\
I_3&=(-\beta+2,0), \\
I_4&=(0,\beta-2),\\
I_5&=(\beta-2,\frac{\beta}{2}),\\
I_6&=(\frac{\beta}{2},\beta-1), \\
I_7&=(\beta-1,-\beta+4), \\
I_8&=(-\beta+4,2),\\
I_9&=(2,\beta).
\end{array}\end{equation}
The inverses or preimages of these intervals are

\begin{equation}\begin{array}{rl}
T^{-1}(I_1) &=I_9,\\
T^{-1}(I_2) &=(\beta-1,2),\\
T^{-1}(I_3) &= (-\beta,-2)\cup I_6,\\
T^{-1}(I_4) &= (-2,\beta-4)\cup (1,\frac{\beta}{2}),\\
T^{-1}(I_5) &= (\beta-4,\frac{\beta-4}{2})\cup (\frac{\beta}{4},1),\\
T^{-1}(I_6) &= (\frac{\beta-4}{2},\beta-3)\cup (\frac{1}{2},\frac{\beta}{4}),\\
T^{-1}(I_7) &= (\beta-3,-\beta+2)\cup (\beta-2,\frac{1}{2}),\\
T^{-1}(I_8) &=(-\beta+2,0)\cup (\frac{\beta-2}{2},\beta-2) ,\\
T^{-1}(I_9) &=(0,\frac{\beta-2}{2}).
\end{array}\end{equation}
We assume the following ansatz for the invariant density

\begin{equation}
\rho (x)= A_i,\quad {\rm if}~x\in I_i,\ \ i=1,\dots ,9, \quad {\rm with} \quad A_7=A_8
\end{equation}
Considering (\ref{eq:PFbeta2}) this gives the equations

\begin{equation}\begin{array}{rl}
A_1&=\frac{1}{2}A_9,\\
A_2&=\frac{1}{2}A_7,\\
A_3&=A_1+\frac{1}{2}A_6,\\
A_4&=A_1+\frac{1}{2}A_5,\\
A_5&=A_2+\frac{1}{2}A_5,\\
A_6&=A_2+\frac{1}{2}A_5,\\
A_7&=A_2+\frac{1}{2}A_5,\\
A_8&=A_3+\frac{1}{2}A_4,\\
A_9&=\frac{1}{2}A_4.\\
\end{array}\end{equation}

It is already clear from these equations that $A_5=A_6=A_7=A_8$ and $A_3=A_4$. The additional normalization requirement reads
\begin{eqnarray}
&&(2\beta-4)A_1+(6-2\beta)A_2+(\beta-2)A_3+(\beta-2)A_4\nonumber\\
&+&(2-\frac{\beta}{2})A_5
+(\frac{\beta}{2}-1)A_6+(5-2\beta)A_7+(\beta-2)A_8\nonumber\\
&+&(\beta-2)A_9=1.
\end{eqnarray}
From here it is trivial to obtain the solution

\begin{equation}\begin{array}{rl}
A_1&= \textstyle{\frac{1}{4}}A_4, \\
A_2&=\textstyle{\frac{3}{4}}A_4,  \\
A_3&=A_4,\\
A_4&=\frac{2}{9},\\
A_5&=A_6=A_7=A_8=\textstyle{\frac{3}{2}}A_4, \\
A_9&=\textstyle{\frac{1}{2}}A_4,
\end{array}\end{equation}
which is independent of $\beta$. Finally, the measure of the interval $[0,\beta]$
\begin{eqnarray}
\mu ([0,\beta])&=&(\beta-2)A_4+(2-\frac{\beta}{2})A_5+(\frac{\beta}{2}-1)A_6\nonumber\\
&&+(5-2\beta)A_7+
(\beta-2)A_8+(\beta-2)A_9\nonumber\\
&=&(\beta-2)A_4+(4-\beta)A_8+(\beta-2)A_9\nonumber\\
&=&[(\beta-2)+\frac{3}{2}(4-\beta)+\frac{1}{2}(\beta-2)]A_4\nonumber\\
&=&\frac{2}{3},
\end{eqnarray}
which is also independent of $\beta$ and produces therefore a constant Lyapunov exponent.

Using the same method it is straightforward to prove that the invariant density is the same for the whole range $2<\beta<3$. Therefore, we can write

\begin{equation}
 \rho_{\beta\in(2,3)} (x) =\left\{\begin{array}{ll}
\frac{1}{18} & {\rm if}~x\in (-\beta , \beta-4)\\
\frac{1}{6} & {\rm if}~x\in (\beta-4 , -\beta+2)\\
\frac{2}{9} & {\rm if}~x\in (-\beta+2, \beta-2)\\
\frac{1}{3} & {\rm if}~x\in (\beta-2 , 2)\\
\frac{1}{9} & {\rm if}~x\in (2 , \beta)
 \end{array} .\right.
\end{equation}
We can check that the measure $\mu ([0,\beta])$ for this invariant density is again $\frac{2}{3}$
\begin{eqnarray}
\mu_{\beta\in(2,3)} ([0,\beta])=(\beta-2)\frac{2}{9}+(4-\beta)\frac{1}{3}+(\beta-2)\frac{1}{9}=\frac{2}{3}.
\end{eqnarray}
By looking at the bifurcation diagram in figure~\ref{fig:LyapBif} it is possible to see that the invariant density changes at $\beta=3$. At this particular value the density is

\begin{equation}
 \rho_{\beta=3} (x) =\left\{\begin{array}{ll}
\frac{1}{18} & {\rm if}~x\in (-3, -1),\\
\frac{2}{9} & {\rm if}~x\in (-1 , 1),\\
\frac{1}{3} & {\rm if}~x\in (1, 2),\\
\frac{1}{9} & {\rm if}~x\in (2 , 3),
 \end{array} \right.
\end{equation}
and again $\mu_{\beta=3} ([0,\beta])=\frac{2}{9}+\frac{1}{3}+\frac{1}{9}=\frac{2}{3}$.

In the range $3<\beta<4$ the density is given by

\begin{equation}
 \rho_{\beta\in(3,4)} (x) =\left\{\begin{array}{ll}
\frac{1}{18} & {\rm if}~x\in (-\beta ,- \beta+2),\\
\frac{1}{9} & {\rm if}~x\in (-\beta+2 , \beta-4),\\
\frac{2}{9} & {\rm if}~x\in (\beta-4, \beta-2),\\
\frac{1}{3} & {\rm if}~x\in (\beta-2 , 2),\\
\frac{1}{9} & {\rm if}~x\in (2 , \beta).
 \end{array}\right.
\end{equation}

The point $\beta=4$ is again an special point where the density changes. At this particular point the density is given by

\begin{equation}
 \rho_{\beta=4} (x) =\left\{\begin{array}{ll}
\frac{1}{18} & {\rm if}~x\in (-4, -2),\\
\frac{1}{9} & {\rm if}~x\in (-2 , 0),\\
\frac{2}{9} & {\rm if}~x\in (0, 2),\\
\frac{1}{9} & {\rm if}~x\in (2 , 4).
 \end{array}\right.
\end{equation}

Finally, the density in the range $4<\beta<\frac{9}{2}$ is
\begin{equation}
 \rho_{\beta\in(4,\frac{9}{2})} (x) =\left\{\begin{array}{ll}
\frac{1}{27} & {\rm if}~x\in (-\beta , 3\beta-16),\\
\frac{1}{18} & {\rm if}~x\in (3\beta-16 , -\beta+2),\\
\frac{5}{54} & {\rm if}~x\in (-\beta+2, 3\beta-14),\\
\frac{1}{9} & {\rm if}~x\in (3\beta-14 , -\beta+4),\\
\frac{4}{27} & {\rm if}~x\in (-\beta+4 , \beta-4),\\
\frac{2}{9} & {\rm if}~x\in (\beta-4 , -\beta+6),\\
\frac{7}{27} & {\rm if}~x\in (-\beta+6 , 2),\\
\frac{1}{9} & {\rm if}~x\in (2 , -\beta+8),\\
\frac{2}{27} & {\rm if}~x\in (-\beta+8 , \beta).\\
 \end{array}\right.
\end{equation}

It is straightforward to check that the invariant measure $\mu([0,\beta])$ for all the densities $\rho(x)$ we have just defined is $\frac{2}{3}$. Therefore we have been able to prove the constancy of the Lyapunov exponent for the case $s=2$ in the range $2<\beta<\frac{9}{2}$. The Lyapunov exponent takes the constant value
\begin{equation}
\lambda=\frac{2}{3}\ln 2.
\end{equation}

For $\beta>\frac{9}{2}$ the number of pieces of the invariant density increases considerably and its calculation becomes much more involved. In particular, it is very difficult to find an appropriate partition on which we can define an ansatz of $\rho(x)$ that solves the Perron-Frobenius equation.

\subsection{$\beta<2$}

Using the same method it is also possible to study the dependence on $\beta$ of the Lyapunov exponent just before the start of the plateau. In this case the Perron-Frobenius equation reads
\begin{equation}\label{eq:PFbeta3}
 \rho (x) =\left\{\begin{array}{ll}
 \frac{1}{2}\rho (\frac{1}{2}(\beta -x)) & {\rm if}~x\in (\beta-4 , \beta-2),\\
 \rho (x-2)+ \frac{1}{2}\rho (\frac{1}{2}(\beta -x)) & {\rm if}~x\in (\beta-2 , 2),\\
 \end{array}\right.
\end{equation}
Making use of the following ansatz
\begin{equation}
 \rho (x) =\left\{\begin{array}{ll}
 A & {\rm if}~x\in (\beta-4 ,\beta-2),\\
 B  & {\rm if}~x\in (\beta-2 , 2),\\
 \end{array}\right.
\end{equation}
and taking into account the normalization we get

\begin{equation}\label{eq:PFbeta2}
 \rho_{\beta<2} (x) =\left\{\begin{array}{ll}
 \frac{1}{2(5-\beta)} & {\rm if}~x\in (\beta-4 , \beta-2),\\
 \frac{1}{5-\beta}  & {\rm if}~x\in (\beta-2 , 2).\\
 \end{array}\right.
\end{equation}
This results in the following Lyapunov exponent for $\beta<2$

\begin{equation}
\lambda=\frac{2}{5-\beta}\ln 2.
\end{equation}

\section{Lyapunov exponent plateau for $s=\phi=\frac{1+\sqrt{5}}{2}$}
\label{app:proofsGR}

\subsection{$2\phi<\beta<4$}
Similarly to the case studied in the previous appendix, the Perron-Frobenius equation (\ref{eq:PF}) is given by

\begin{equation}\label{eq:PFbetaGR}
 \rho (x) =\left\{\begin{array}{ll}
 \frac{1}{\phi}\rho (\frac{1}{\phi}(\beta -x)) & {\rm if}~x\in (\beta(1-\phi) , \beta(1-\phi)+2),\\
 \rho (x-2)+ \frac{1}{\phi}\rho (\frac{1}{\phi}(\beta -x)) & {\rm if}~x\in (\beta(1-\phi)+2 , 2),\\
 \frac{1}{\phi}\rho (\frac{1}{\phi}(\beta -x)) & {\rm if}~x\in (2 , \beta).
 \end{array}\right.
\end{equation}
We consider now a partition in eight intervals

\begin{equation}\begin{array}{rl}
I_1&=(\beta(1-\phi),2(\beta-2\phi-1)), \\
I_2&=(2(\beta-2\phi-1),\beta(1-\phi)+2), \\
I_3&=(\beta(1-\phi)+2,\beta-2\phi), \\
I_4&=(\beta-2\phi,2\beta-4\phi), \\
I_5&=(2\beta-4\phi,\beta(1-\phi)+4), \\
I_6&=(\beta(1-\phi)+4,2), \\
I_7&=(2,\beta(1-\phi)+2\phi^2), \\
I_8&=(\beta(1-\phi)+2\phi^2,\beta).
\end{array}\end{equation}
The preimages of these intervals are

\begin{equation}\begin{array}{rl}
T^{-1}(I_1) &=I_8,\\
T^{-1}(I_2) &=(\frac{1}{\phi}(\beta\phi-2),\beta(1-\phi)+2\phi^2),\\
T^{-1}(I_3) &= (\beta(1-\phi),\beta-2\phi-2)\cup (2,\frac{1}{\phi}(\beta\phi-2)),\\
T^{-1}(I_4) &= (\beta-2\phi-2,2(\beta-2\phi-1))\cup I_6,\\
T^{-1}(I_5) &= I_2\cup (\beta+4(1-\phi),\beta(1-\phi)+4),\\
T^{-1}(I_6) &= (\beta(1-\phi)+2, 0)\cup (\frac{\beta-2}{\phi},\beta+4(1-\phi)),\\
T^{-1}(I_7) &= (\beta-2\phi,\frac{\beta-2}{\phi}),\\
T^{-1}(I_8) &=(0,\beta-2\phi).
\end{array}\end{equation}

Assuming the following ansatz for the invariant density

\begin{equation}
\rho (x)= A_i,\quad {\rm if}~x\in I_i,\ \ i=1,\dots ,8, \quad {\rm with} \quad A_4=A_5.
\end{equation}
(\ref{eq:PFbetaGR}) gives the equations
\begin{equation}\begin{array}{rl}
A_1&=\frac{1}{\phi}A_8,\\
A_2&=\frac{1}{\phi}A_7,\\
A_3&=A_1+\frac{1}{\phi}A_7,\\
A_4&=A_1+\frac{1}{\phi}A_6,\\
A_5&=A_2+\frac{1}{\phi}A_5,\\
A_6&=A_3+\frac{1}{\phi}A_5,\\
A_7&=\frac{1}{\phi}A_5,\\
A_8&=\frac{1}{\phi}A_3.\\
\end{array}\end{equation}
The additional normalization requirement reads
\begin{eqnarray}
&&(\beta(1+\phi)-2(1+2\phi))A_1+(4-\beta)(\phi+1)A_2\nonumber\\
&+&((\beta-2)\phi-2)A_3+(\beta-2\phi)A_4+(4-\beta)(\phi+1)A_5\nonumber\\
&+&(-2-\beta(1-\phi))A_6+(\beta(1-\phi)+2\phi)A_7\nonumber\\
&+&((\beta-2)\phi-2)A_8+(\beta-2)A_9=1.
\end{eqnarray}
From here it is trivial to obtain the solution

\begin{equation}\begin{array}{rl}
A_1&= \textstyle{\frac{1}{\phi^3}}A_5,\\
A_2&=\textstyle{\frac{1}{\phi^2}}A_5, \\
A_3&=\textstyle{\frac{1}{\phi}}A_5,\\
A_4&=A_5,\\
A_5&=\frac{1}{10-4\phi}\\
A_6&=\textstyle{(\frac{1}{\phi^3}+1)}A_5, \\
A_7&=\textstyle{\frac{1}{\phi}}A_5, \\
A_8&=\textstyle{\frac{1}{\phi^2}}A_5,
\end{array}\end{equation}
which is independent of $\beta$. Finally, the measure of the interval $[0,\beta]$ is

\begin{equation}
\mu ([0,\beta])=\frac{3-\phi}{5-2\phi},
\end{equation}
which is also independent of $\beta$ and results in the constant Lyapunov exponent
\begin{equation}
\lambda=\frac{3-\phi}{5-2\phi}\ln\phi
\end{equation}

\subsection{$\beta<2\phi$ }

To obtain the analytical form of the Lyapunov exponent before the plateau, we consider the same Perron-Frobenius equation (\ref{eq:PFbetaGR}). However, now we divide the phase space in five intervals
\begin{equation}\begin{array}{rl}
I_1&=(\beta(1-\phi),\beta-2\phi), \\
I_2&=(\beta-2\phi,\beta(1-\phi)+2), \\
I_3&=(\beta(1-\phi)+2,\beta-2\phi+2), \\
I_4&=(\beta-2\phi+2, 2),\\
I_5&=(2, \beta),
\end{array}\end{equation}
with inverses
\begin{equation}\begin{array}{rl}
T^{-1}(I_1) &=I_5,\\
T^{-1}(I_2) &=I_4,\\
T^{-1}(I_3) &= I_1\cup (2(2-\phi),\beta-2\phi+2),\\
T^{-1}(I_4) &= (\beta-2\phi,0)\cup (\frac{\beta-2}{\phi},2(2-\phi)),\\
T^{-1}(I_5) &= (0,\frac{\beta-2}{\phi}).
\end{array}\end{equation}

Using the following ansatz for the invariant density
\begin{equation}
\rho (x)= A_i,\quad {\rm if}~x\in I_i,\ \ i=1,\dots ,5, \quad {\rm with} \quad A_2=A_3
\end{equation}
the Perron-Frobenius equation produces the following set of equations
\begin{equation}\begin{array}{rl}
A_1&=\frac{1}{\phi}A_5,\\
A_2&=\frac{1}{\phi}A_4,\\
A_3&=A_1+\frac{1}{\phi}A_3,\\
A_4&=A_2+\frac{1}{\phi}A_3,\\
A_5&=\frac{1}{\phi}A_3.
\end{array}\end{equation}

Considering the normalization it is straightforward to find
\begin{equation}\begin{array}{rl}
A_1&= \textstyle{\frac{1}{\phi^2}}A_3,\\
A_2&=A_3, \\
A_3&=\frac{1}{\beta\phi-2(\beta+\phi-4)},\\
A_4&=\phi A_3,\\
A_5&=\textstyle{\frac{1}{\phi}}A_3.
\end{array}\end{equation}
The measure of the interval $[0,\beta]$ is therefore

\begin{equation}
\mu ([0,\beta])=2(3-\phi)A_3,
\end{equation}
and the Lyapunov exponent is

\begin{equation}
\lambda=\frac{2(3-\phi)}{\beta\phi-2(\beta+\phi-4)}\ln\phi.
\end{equation}

\section*{References}
\bibliographystyle{unsrt.bst}
\bibliography{Plateau_ms.bib}

\end{document}